\documentclass{agujournal2019}
\usepackage{url} 
\usepackage{lineno}
\usepackage[inline]{trackchanges} 
\usepackage{soul}
\usepackage{xcolor,amsmath}
\usepackage{adjustbox}

\usepackage{rotating} 
\usepackage[nomarkers,nolists]{endfloat}  
\DeclareDelayedFloatFlavor{sidewaystable}{table} 
\DeclareDelayedFloatFlavor{sidewaystable*}{table}

\draftfalse

\journalname{Journal of Advances in Modeling Earth Systems (JAMES)}

\begin{document}

%
%


\title{
Regional Climate Model Emulation with Diffusion Approaches: What is the Added Value of Generative Machine Learning?}

%
%




\authors{Mikel N. Legasa\affil{1,2}, Antoine Doury\affil{3}, Achille Gellens\affil{1,2}, Redouane Lguensat\affil{2}, Clara Naldesi\affil{1,2,4}, Soulivanh Thao\affil{1,2}, Mathieu Vrac\affil{1,2}}

\affiliation{1}{Laboratoire des Sciences du Climat et de l’Environnement (LSCE), CEA/CNRS/UVSQ, Université Paris Saclay, UMR8212, 91191 Gif-sur-Yvette, France}
\affiliation{2}{Institut Pierre-Simon Laplace (IPSL), FR636, Paris, France}
\affiliation{3}{CNRM, Université de Toulouse, Météo-France, CNRS, Toulouse, France}
\affiliation{4}{Autorité de sûreté nucléaire et de radioprotection, PSE-ENV/SCAN/BEHRIG, F-92260, Fontenay-aux-Roses, France}




\correspondingauthor{Mikel N. Legasa}{mikel.legasa@lsce.ipsl.fr}



\begin{keypoints}
\item We introduce ParamDiffusion, a cost-efficient two-stage diffusion-based framework for regional climate model emulation.

\item We propose a novel target-based validation framework, with extreme events, to intercompare four emulators, including two diffusion models.

\item Diffusion models excel at capturing climatological statistics, but their uncertainty envelopes fail to capture some extreme rainfall events.
\end{keypoints}

%
%

%
%


\begin{abstract}
 Emulators provide a cost-effective alternative to regional climate models (RCMs) by statistically capturing their dynamical downscaling function. They link large-scale predictors simulated by global climate models (GCMs) to RCM-simulated high-resolution fields of the target variable, here precipitation. Machine learning methods, typically deep learning, are cheaper than running RCMs in computation time and energy. Among them, generative models are appealing because they can simulate ensembles of local high-resolution fields consistent with the predictors. This ensemble, which we call the \textit{uncertainty envelope}, remains to be properly assessed for added value. Here, we make three contributions. First, we introduce ParamDiffusion, a new two-stage diffusion-based framework, and compare it with a state-of-the-art diffusion approach. Second, we expand standard validation through a comprehensive framework aligned with climate-science needs, examining specific precipitation events, including extremes. Third, within this framework, we assess the added value of diffusion approaches relative to deterministic methods. We intercompare four deep-learning models: a deterministic model designed to capture the precipitation tail; a parametric probabilistic model based on it; a recently proposed diffusion approach; and ParamDiffusion, which couples the parametric model with a diffusion model. Our results show that diffusion-based approaches reproduce climatological precipitation statistics with high skill, including distributional tails and spatially compounded extremes, while generating spatially detailed fields. However, none of the assessed models consistently accounts for the most extreme RCM-simulated events within its uncertainty envelope. Diffusion models are therefore promising for probabilistic RCM emulation, but substantial progress is still required before they can reliably represent high-impact precipitation extremes.
\end{abstract}

\section*{Plain Language Summary}

Regional climate models are used to simulate local climate variables such as precipitation at high resolutions, but they are expensive to run. A cheaper alternative is to train machine learning/artificial intelligence \textit{emulators} to reproduce their precipitation fields from large-scale atmospheric conditions. In this study, we assess whether a class of generative artificial intelligence models, called diffusion models, can provide not only realistic precipitation maps, but also a useful range of possible precipitation patterns for a given atmospheric situation, thus allowing for the exploration of uncertainty. We also introduce ParamDiffusion, a new approach, and compare it with a direct diffusion model, a simpler probabilistic model, and a model designed specifically for precipitation that produces only one precipitation map. We evaluate the models using long-term precipitation statistics and a detailed analysis of selected precipitation events, including severe extremes. Diffusion models reproduce long-term precipitation statistics well and generate spatially detailed precipitation fields. They also often improve over the single-map emulator for individual events. However, the most intense and localized precipitation extremes are not always well captured. Diffusion models are therefore promising tools for fast climate-model emulation with uncertainty, but more work is needed to capture the most severe precipitation extremes.

%
%

%


%
%
%
%

\section{Introduction}

Global climate models (GCMs), the primary tools for studying future climates, require downscaling to bridge their coarse resolution to local scales. By employing machine learning, \textit{emulators} \cite{chadwick_artificial_2011, babaousmail_novel_2021, boe_simple_2023, doury_regional_2023, rampal_extrapolation_2024} aim to provide a cost-effective alternative to regional climate models (RCMs). They do so by statistically replicating their dynamical downscaling function, therefore constituting a so-called \textit{hybrid} methodology. Emulators work by empirically linking a set of large-scale predictors, well simulated by GCMs, to RCM-simulated high-resolution fields of the variable of interest, such as precipitation, the focus of this work. Downscaling precipitation poses notable difficulties: it is intermittent, positively skewed, spatially complex, and strongly influenced by many different local and large-scale processes \cite{vrac_stochastic_2007, maraun_precipitation_2010, vandal_deepsd_2017, doury_suitability_2024}. In this study we build and evaluate the RCM-emulators in the \textit{perfect model} framework meaning that  both predictors and targets come from the same RCM. Even in this context, the large-scale predictors provide insufficient information to fully determine the high-resolution precipitation fields. From the emulator’s perspective, the downscaling function is therefore intrinsically uncertain  \cite{rampal_downscaling_2025}. Capturing this uncertainty is particularly important under a changing climate, where local precipitation extremes are among the most relevant events to society and ecosystems \cite{clarke_extreme_2022} but also challenging to predict and treat statistically. 

This uncertainty highlights the limits of deterministic emulators, that is, methods that predict a single precipitation field from the large-scale predictors. Over the last few years, deep learning methods have shown considerable skill for climate downscaling and RCM emulation \cite{vandal_deepsd_2017, vaughan_convolutional_2021, wang_fast_2021, doury_regional_2023, doury_suitability_2024}. Deterministic models necessarily produce a single field for each large-scale state, and therefore must make a compromise, producing \textit{smoothed} precipitation fields, and overall under representing extremes. This is especially the case when emulators are trained to minimize loss functions not well suited to precipitation, such as the mean squared error \cite{doury_suitability_2024, tomasi_can_2025, addison_machine_2026}. Loss functions designed specifically for precipitation, two of them explored in the present article, partly address this problem by better accounting for the shape of its distribution. However, the problem remains that deterministic models are unable to capture the uncertainty of the downscaling function.

This makes generative machine learning models a sensible alternative to tackle the problem, since they can represent the conditional distribution of high-resolution fields rather than collapse them into a single best point estimate. This has made them gain much attention in recent years \cite{mardani_residual_2025, rampal_downscaling_2025, schillinger_enscale_2025, tomasi_can_2025, addison_machine_2026}. Among them, \textit{diffusion} models have become a prominent modeling framework. Denoising diffusion probabilistic models \cite[DDPM]{song_scorebased_2021} are a generative machine-learning framework to learn the probability distribution of the data by progressively corrupting samples with Gaussian noise. A deep learning model, typically a U-Net \cite{ronneberger_unet_2015}, acts as a \textit{backbone} and is trained to learn to reverse this process. This backbone can subsequently be used for the generation of new samples via iterative denoising. Note that, throughout this article, we use the term \textit{diffusion} as an umbrella term for this broader family of iterative generative models stemming from DDPM, including recent variants that differ in their precise training objective, stochastic process, or sampling formulation. They can be conditioned on large-scale predictors to act as RCM-emulators and are appealing for downscaling: they produce highly spatially-detailed samples that represent probability distributions rather than single deterministic fields, thus enabling large ensembles of downscaled realizations at substantially lower costs than those required to run an RCM. However, they are still more computationally demanding than state-of-the-art deterministic machine learning approaches, and the extent of their added value when downscaling GCM simulations still remains an open question.

In recent studies, diffusion models have started to demonstrate a lot of potential in weather and climate applications. \citeA{addison_machine_2026} introduced a conditional diffusion emulator for daily precipitation from convection-permitting simulations, showing realistic spatial structures and skill when simulating extremes. \citeA{tomasi_can_2025} proposed to use a latent diffusion model for kilometer-scale downscaling over Italy, for wind and temperature. \citeA{mardani_residual_2025} introduced CorrDiff, a residual corrective diffusion framework with a deterministic model to predict the expected value and a diffusion model to learn the residuals around it, finding promising results. Other studies have proposed diffusion frameworks or other generative methods such as GANs (generative adversarial networks) for emulation and downscaling \cite{aich_conditional_2024, hess_fast_2024, bassetti_diffesm_2024, rampal_extrapolation_2024,  schillinger_enscale_2025}. Together, these studies show how generative methods are gaining widespread attention in the context of climate downscaling and their potential.

What remains less settled is how these models and their added value with respect to deterministic references should be evaluated, especially for precipitation. Existing studies often demonstrate improvements in visual realism, spatial detail, climatological distributions, or selected extreme summaries \cite{mardani_residual_2025, tomasi_can_2025, addison_machine_2026}. While these diagnostics are valuable, they do not fully answer whether the conditional stochastic ensemble produced by the diffusion model is an informative uncertainty envelope given the large-scale conditions. This distinction matters, since a generative model can faithfully reproduce climatological tails and realistic-looking fields but may fail to assign probabilities correctly for specific large-scale conditions, which goes both ways: they may hallucinate realities not really possible, or not fully simulate extremes when these are consistent with the large-scale predictors' state. We believe, thus, that the central question when assessing generative emulators is not whether or not samples look realistic, but whether their conditional distribution is informative for climate analysis. In particular, one of the most interesting features of a generative model is their potential to generate climate extremes consistent with the large-scale predictors' state. For such cases, deterministic models have to compromise on a specific fixed field, which, by definition, will not be an extreme. In this regard, however, the choice of deterministic reference is also important, with studies such as \citeA{addison_machine_2026} having only compared their diffusion approach against a deterministic U-Net trained with mean squared error as loss function. This has been clearly shown to perform poorly \cite{doury_suitability_2024}, and thus we believe the added value of diffusion models may be overstated. If this is the case, the relevant question becomes sharper: do diffusion models add value beyond a competitive deterministic model, and if so, what is the extent of the added value? 

This study addresses this question, over mainland France and northeast Spain and under the perfect model framework (both predictors and predictands come from the same RCM simulation), following \citeA{doury_suitability_2024}. Our contribution with this article is both on the diffusion methodology and on the evaluation. First, we propose a two-step diffusion model methodology that relies on a high-resolution representation of precipitation fields together with their uncertainty, expressed in a Bernoulli-Gamma distribution. This modeling approach is inspired by other two-step approaches such as the one in \citeA{mardani_residual_2025}, and completely separates the deterministic information in the large-scale predictors and the uncertainty envelope. This results in a substantially computationally cheaper as well as potentially more transferable approach. With respect to the evaluation, on top of traditional evaluation methods such as power spectral density or climatologies, we assess how informative the uncertainty cloud generated by diffusion-based RCM emulators actually is. More specifically, this study aims to assess whether diffusion approaches provide added value by taking into account specific precipitation targets of interest to the climate community, such as extremes. This evaluation is performed by using a state-of-the-art deterministic precipitation model adapted to capture the tail of the precipitation distribution, a parametric model directly based on it, and two diffusion models: our proposed approach, ParamDiffusion, as well as the state-of-the-art  diffusion model proposed in \citeA{addison_machine_2026}. 

By doing so, we aim to provide not only a comprehensive model intercomparison, but also move forward towards a more comprehensive validation framework for generative climate emulation.

\section{Experimental Framework and Methods}

\subsection{Experimental Framework}\label{s.experimental_framework}

The experimental framework adopted in this study follows that of \citeA{doury_suitability_2024}. The RCM emulated is ALADIN63 \cite{nabat_modulation_2020}, driven by CNRM-CM5 (run r1i1p1), within the EURO-CORDEX framework \cite{coppola_assessment_2021}, at a spatial resolution of around 12.5 km. The domain considered for this study is composed of 64x64 gridpoints
 covering most of mainland France and a small part of northeast Spain (4º West - 4º East, 42º North - 50º North,  see Figure \ref{f:targets}). We work within a \textit{perfect-model} setting, in which both predictors and target fields are derived from the same RCM simulation. The large-scale predictors (geopotential, humidity, temperature, eastward and northward wind components at 850, 700, 500 hPa level pressures, as well as sea level pressure) are obtained by conservatively coarsening the RCM fields to a typical GCM resolution (around 150 km, covering the domain -10º West - 20º East, 36º North - 57º North), following \citeA{doury_suitability_2024}, and are then used as input to emulate the corresponding high-resolution RCM precipitation field. This strategy isolates the RCM downscaling function from additional sources of uncertainty associated with GCM--RCM inconsistencies, which may otherwise limit transferability in so-called \textit{imperfect} frameworks  \cite{bano-medina_transferability_2023, boe_simple_2023, vandermeer_deep_2023}. It also provides us a controlled benchmark in which we consider the RCM-simulated precipitation fields as the reference \textit{truth} that the emulators should reproduce.

Importantly, this setup does not imply that the downscaling relationship is deterministic from the emulator’s perspective: although an RCM produces a single high-resolution realization for a given large-scale atmospheric state, the predictors contain incomplete information about local precipitation at higher resolution. State-of-the-art deterministic emulators therefore provide a useful point estimate, but their skill remains limited, with an average spatial  spearman correlation of around 0.83 after removing the seasonal cycle (for our current domain), leaving some unresolved variability. The goal of using a diffusion model is precisely to characterize this conditional uncertainty envelope around the expected high-resolution field, rather than improving the mean prediction.

This study focuses on the evaluation within the perfect model framework, that is, in \textit{perfect conditions}. Once trained and in final application, emulators are meant to be applied to the GCM-simulated large-scale variables; however, in this study, we are only assessing the capability of models to perfectly replicate the downscaling function. Throughout the whole article, models are trained with data from the Historical plus RCP8.5 scenarios, thus the combined years from 1951 to 2100. Subsequently, the trained models are applied to predictors from the RCP4.5 scenario, for the period 2005-2100. All the evaluations and results shown in this article, in Section \ref{s:results}, correspond to simulations and predictions performed for the evaluation period. This framework tests the models on unseen data and, although it is not the main objective of this article, it also explores transferability to other scenarios.

\subsection{Models}

We list in this section the four deep learning methods considered in this work, with a subsection dedicated to each of them. They are also summarized in Table \ref{t:models}. Note that all of them have been trained with Adam optimizer \cite{kingma_adam_2015} keeping $10\%$ of the training data as validation set to early stop the algorithm once no further improvement was observed for the specific loss function of each method.

\begin{sidewaystable}
\centering
\begin{tabular}{ |c|l|c|c|c|c| }
\hline
 \textbf{Name} & \textbf{Description} & \textbf{Parameters}  & \textbf{Type} & \textbf{Training Time} & \textbf{Simulation Time} \\ \hline
 Asym & U-Net with loss function to capture the tail. & 389 471& Deterministic & 6 h & $<$ 10 s  \\  
 B-Gamma & U-Net with gridpoint-based parametric output & 389 533& Parametric & 6 h & $<$ 1 m \\
 ParamDiffusion & Diffusion on B-Gamma output. & $\sim$ 0.4 + 15.7 M & Generative & 6 + 14 h  & 16 h \\ 
 CPMGEM & Diffusion on the large-scale predictors. & $\sim$  63 M & Generative & 63 h  & 48 h \\ \hline
\end{tabular}
\vspace{0.2cm}
\caption{The different methodologies assessed in this article. Note the number of parameters of the ParamDiffusion model is shown as a sum for the background model + the diffusion backbone  (the diffusion backbone has 15 724 929 parameters). A summary of computational costs is provided in the last two columns. For the four models, the times are reported for the same GPU model: a \textit{Nvidia Tesla V100-SXM2-16GB} (both training and simulating). Simulation time corresponds to the total time required to produce one simulation (prediction in the case of Asym) for the whole test period considered in this study  (RCP4.5, 2005-2100). Numbers are rounded to the nearest unit. Training times for ParamDiffusion are shown as a sum (background + diffusion backbone).} \label{t:models}
\end{sidewaystable}

\subsubsection{Asym: Determinist U-Net Model for Precipitation}
 \citeA{doury_regional_2023} developed a deep learning RCM emulator specifically designed to capture the RCM downscaling process of temperature, employing the well-known U-Net architecture \cite{ronneberger_unet_2015}, essentially an encoder–decoder architecture with skip connections. Subsequently, in \citeA{doury_suitability_2024} this model was upgraded to downscale precipitation fields, mainly by updating the loss function for the semicontinuous and highly skewed distribution of precipitation.  Beyond the spatial atmospheric predictors, the architecture includes a dense neural network that processes a series of non-spatial predictors, such as seasonal indicators, external forcings, and spatial summary statistics of the input fields. The dense-neural network output is concatenated with the spatial representation at the bottleneck of the U-Net. We consider this model the state-of-the-art deterministic deep learning model for RCM emulation of precipitation.
 
Note that the original model contains approximately 28 million trainable parameters. In this work, we found the exact configuration in \citeA{doury_suitability_2024} to be suboptimal and overly complex for our specific experimental framework and thus we have performed a hyperparameter optimization. Qualitatively, the model remains the same, with the following changes: instead of the $4$ contractive and expansive blocks (with $64$, $128$, $256$, $512$ kernels) used in the original article; we use only $2$ blocks with $32$ and $64$ kernels. The scalar predictors' processing block also uses $2$ dense layers with $16$, $32$ neurons, instead of the $4$ (with $64$, $128$, $256$, $512$ neurons) in the original article. This results in an architecture with a substantially lower number of parameters (around $390 000$ trainable parameters, thus around $\sim 1.4 \%$ the size of the original model), and yields slightly better results (the comparison is not shown). We refer the reader to \citeA{doury_suitability_2024} for further details.

\subsubsection{B-Gamma: Parametric Bernoulli-Gamma U-Net}
Following from the previous section, we can devise a straightforward probabilistic/generative approach, which we call B-Gamma throughout this article, by upgrading the U-Net to output the parameters of a Bernoulli-Gamma distribution. The idea to output the parameters of a Bernoulli-Gamma mixture distribution with a neural network was proposed by \citeA{cannon_probabilistic_2008}, and has been widely applied for statistical downscaling as output for different machine learning models \cite{carreau_stochastic_2011}, including random forests \cite{legasa_posteriori_2022, legasa_assessing_2023} and convolutional neural networks \cite{bano-medina_configuration_2020}. Our B-Gamma model is a compact U-Net trained to minimize the negative log-likelihood as loss function to output the parameters of a Bernoulli-Gamma distribution for precipitation. It has the exact same architecture as Asym, except for the last layer, which outputs 3 fields: $p$, $\alpha$  and  $\beta$ gridpoint parameters of the Bernoulli and Gamma distributions of precipitation. These parameters can be subsequently used to draw samples from the corresponding Bernoulli-Gamma distribution (refer  to, e.g., \citeA{cannon_probabilistic_2008}, for the details). 

We observe an important issue with this model: simulations from parametric models like this one produce inconsistent \textit{fragmented} spatial patterns, which can clearly be seen in Figure \ref{f:targets} (see the B-Gamma samples). Note that this same modeling framework, pointwise parametric formulation for gridded datasets (no matter the architecture e.g., \citeA{carreau_stochastic_2011, bano-medina_downscaling_2022, rampal_highresolution_2022, soares_highresolution_2024, legasa_strengths_2026}), is poorly suited for spatially coherent emulation and downscaling: although it may reproduce marginal distributions at individual gridpoints, the conditional spatial independence implicit assumption prevents the model from generating physically plausible fields. 

Nevertheless, this model is still useful, since it provides a realistic description of gridpoint-based statistics. In addition, it serves as background model for the diffusion model explained in the next section. The full model description and configuration can be found in the GitHub repository (\url{https://github.com/MNLR/Diffusion}). Note that we refer to this model as \textit{generative} throughout this article when we assess simulations drawn from its predicted distribution.

\subsubsection{ParamDiffusion: Diffusion on the parametric output of B-Gamma}
The model we propose, which we name ParamDiffusion, follows a two-step approach, and was inspired by \textit{residual} diffusion-based methods \cite{mardani_residual_2025}. In essence, the first step involves using B-Gamma, detailed in the previous section, to predict the high-resolution Bernoulli-Gamma fields describing the local precipitation distributions for each day. Subsequently, a diffusion model is conditioned on these parameters to produce spatially realistic precipitation samples. We consider, therefore, that the information in the large-scale predictors on the pointwise precipitation is, as observed and shown later in the results, well captured by B-Gamma. This includes its gridpoint uncertainty, so we task the diffusion model with capturing the spatial distribution together with its uncertainty cloud. Note that, differently from the model proposed in \citeA{mardani_residual_2025} our diffusion model is neither residual nor it uses large-scale information directly, but relies on a high-resolution representation of precipitation fields together with their uncertainty, expressed in the Bernoulli-Gamma distribution. In the remainder of this section, we overview the technical details of the diffusion step of the modeling approach.

The diffusion step consists of a conditional denoising diffusion probabilistic model (DDPM, \citeA{ho_denoising_2020}). During training, Gaussian noise is progressively added to transform high-resolution precipitation fields into a standard normal distribution, while a denoising deep learning model (the \textit{backbone}) is simultaneously trained to reverse the process, conditional on the fields predicted by B-Gamma. The conditioning fields are therefore not the original large-scale atmospheric predictors, but a high-resolution and compact probabilistic representation of precipitation obtained from the Bernoulli-Gamma model. As backbone, we use a U-Net architecture, receiving as conditioning $64 \times 64$ Bernoulli-Gamma fields, with three resolution levels and attention layers at the lowest-resolution stages, with 15 724 929 trainable parameters. The model uses 1000 diffusion steps with a squared-cosine noise schedule, proposed in \citeA{nichol_improved_2021}. At inference time, simulations are generated by starting from Gaussian noise fields and iteratively applying the reverse diffusion process, conditional on the B-Gamma output for the target day. Each reverse trajectory produces one spatially coherent precipitation realization. The full model description and configuration can be found in the GitHub repository (\url{https://github.com/MNLR/Diffusion}). Note that, although reducing the number of steps is an option when doing inference without a substantial loss in performance, we intentionally kept them at 1000 to compare our ParamDiffusion against the diffusion approach explained in the next subsection.

\subsubsection{CPMGEM: Direct Conditional Diffusion}\label{ss:methods_addison}

The second diffusion model approach, called CPMGEM (for Convection-Permitting Model Generative Emulator), was recently proposed for RCM emulation in \citeA{addison_machine_2026}. Although in the present study we are not emulating a convection permitting RCM, we keep this same name throughout the article to avoid confusion. The model is based on the so-called NCSN++ backbone (Noise
Conditional Score Network: a U-Net, see \citeA{ho_denoising_2020}) within a sub-variance-preserving stochastic differential equation formulation \cite{song_scorebased_2021}, and samples are generated by solving the reverse stochastic differential equation using the Euler–Maruyama method \cite{bayram_numerical_2018}. Differently to ParamDiffusion, CPMGEM generates stochastic high-resolution precipitation fields directly conditioned on coarse-resolution atmospheric predictors, and is therefore trained to represent the full downscaling function. To do so, the coarse predictor variables are first regridded to the target grid using nearest-neighbour interpolation and then concatenated with the noisy target precipitation field at each denoising step. We have used this model as is, by running the code provided together with the article by \citeA{addison_machine_2026}. In total, the backbone U-Net for CPMGEM has around 63 million parameters, around 4 times the number of parameters of ParamDiffusion, but uses the same number of denoising steps: 1000. We refer the reader to \citeA{addison_machine_2026} for further details.

Together, ParamDiffusion and CPMGEM represent the two mainly used diffusion-based strategies for RCM emulation, which can be broadly framed as \textit{two-stage} or \textit{one-stage} formulations. ParamDiffusion represents the former: a comparatively simple first-stage model extracts the predictable large-scale component, while the diffusion model is tasked with representing the remaining conditional uncertainty envelope with spatial accuracy. By contrast, CPMGEM follows a one-stage formulation, in which the diffusion model is tasked with capturing the full downscaling function, that is, the complete mapping from large-scale predictors to high-resolution precipitation. Therefore, the comparison in this article assesses not only architectural and computational differences, but also the implications of assigning the full downscaling task to the generative model itself versus employing a background model to first extract all information from the large-scale predictors.

\subsection{Evaluation Methods}\label{ss:evaluation_methods}
We use a comprehensive set of evaluation methods to study how the four models described in the previous section perform. They can be divided in two categories: the climatological metrics, intended as a baseline overall assessment of the models considered, and computed over the whole evaluation period; and the targeted analysis, in which we take 8 specific days of special relevance to climate science and perform an in-depth analysis of the added value of generative methods. The results section (Section \ref{s:results}) is divided accordingly, into sections \ref{s:climatological_analysis} and \ref{s:targeted_analysis}.  The metrics are explained next, and summarized in Table \ref{t:metrics}.

The first set of metrics is standard in literature. First, we evaluate the root mean squared error (RMSE). In the case of the diffusion models, we use their expected value, obtained by averaging 50 simulations. Although this is not the intended use of generative approaches, it provides a basic estimate of the models' capacity to preserve the deterministic component of the downscaling function. We then evaluate relative biases for several representative statistics: the percentage of dry days (Dry\%), and, considering only day–gridpoint pairs with precipitation exceeding 1 mm (hereafter referred to as \textit{wet} day-gridpoints), the mean (Mean), standard deviation (SD), median (Median), and 99th percentile of precipitation (P99). In the case of the generative models, the metrics are the average of 50 simulations: 50 simulations are drawn, a statistic is computed from each, and the expected value for the statistic is shown. 

Traditionally, these metrics have been evaluated per gridpoint. This, however, can potentially hide significant model flaws and offers an incomplete picture of model performance. We propose here to pair gridpoint-per-gridpoint bias analyses with an additional assessment that allows us to address also the spatial performance for each statistic: prior to computing the statistic (e.g., P99) we apply a 3x3 average moving window to both the reference and simulated/predicted field. Subsequently, we compute the metric for the resulting field, that now also encompasses local spatial information due to the local averaging. As we show in the results section, this relatively simple method is a powerful tool that can address each statistic's compounded spatial aspect (e.g., the spatial tail for the 3x3 P99). Note that, the larger the window, the less spatially local the aspect analyzed and, on the limit, it corresponds to domain-wide spatial statistics. We have performed this analysis on other spatial scales (e.g., 16x16) and the results are similar to those obtained using a 3x3 moving window. For studies with bigger domains, we consider that combining different window sizes is a sensible approach to assess biases at different scales. We finalize the climatological analysis with the radially averaged power spectral density (RAPSD, \citeA{harris_multiscale_2001, sinclair_empirical_2005}), which compares the complexity of structures (that is, the variability over different spatial scales) produced by different approaches.

\begin{figure}[tb]
\centering
  \makebox[0pt][c]{%
    \includegraphics[width=1.2\linewidth]{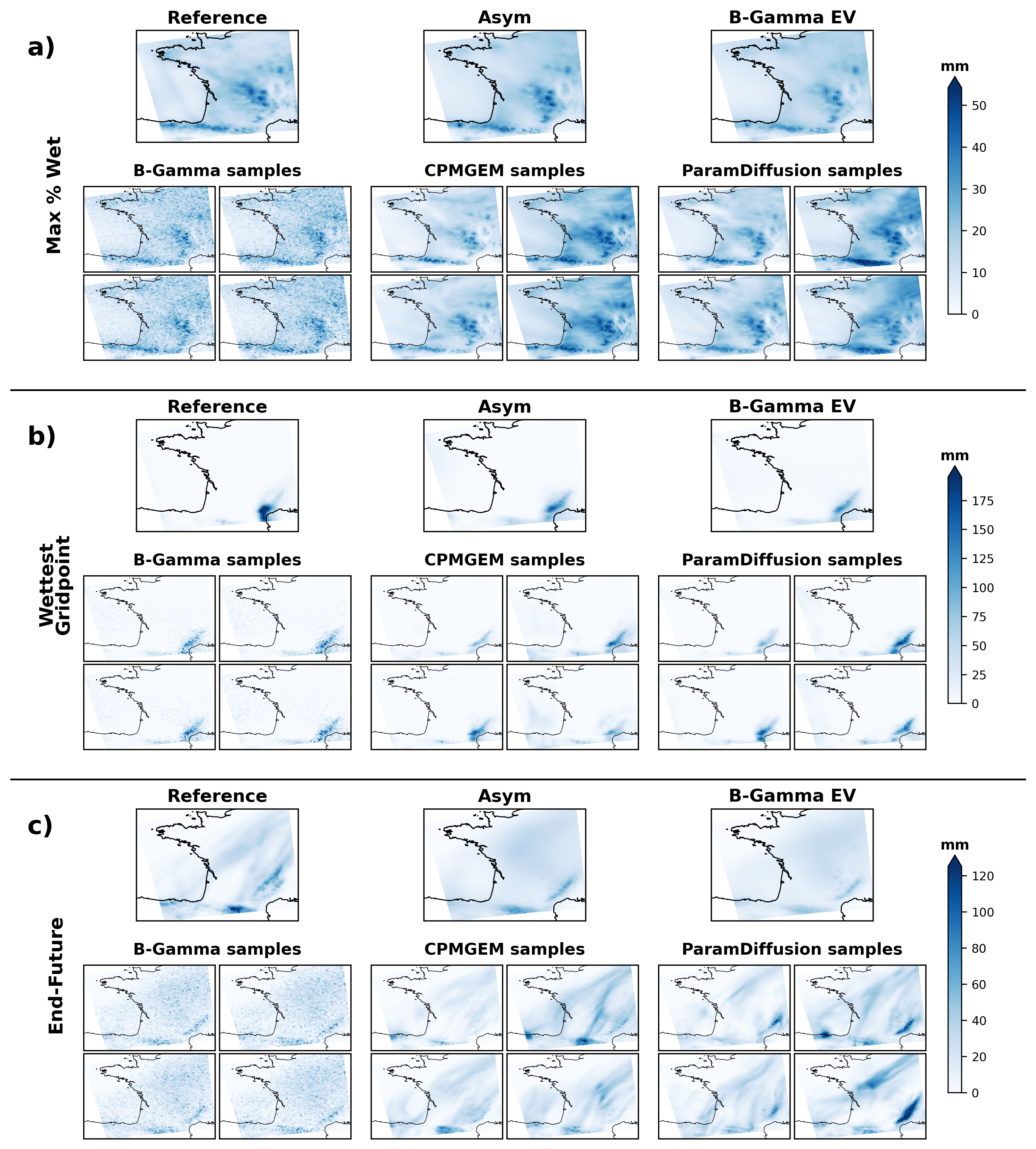}
  }
  \caption{Each panel shows a target selected for the in-depth analysis performed in Section \ref{s:targeted_analysis} (see Table \ref{t:targets}). For each panel, the first row shows the reference precipitation field, as predicted by Asym, and as predicted (expected value) by B-Gamma. The second row shows 4 simulations for each generative model (B-Gamma, CPMGEM, ParamDiffusion). The 4 simulations correspond to, respectively and clockwise: the target with the lowest rainfall amount, the target with the highest rainfall amount, the target closest to the reference (in terms of gridpoint mean absolute error) and the target farthest. }
\label{f:targets}
\end{figure}

The remaining metrics are computed specifically for 8 dates selected from the validation period (RCP 4.5, 2005-2100) and are aimed to evaluate the uncertainty envelope generated by diffusion models. The first five targets correspond different forms of precipitation extremes, which are particularly relevant for assessing whether the models can reproduce situations with substantial potential impacts \cite{clarke_extreme_2022}. These include the day with the highest precipitation aggregated over the whole domain, and the day with the highest precipitation over any gridpoint. The dates for the last three, named \textit{Historical, Mid-future} and \textit{End-Future} were chosen in \citeA{doury_suitability_2024} arbitrarily, and serve us to represent also three standard targets. A detailed description and statistics of the eight target days can be checked in Table \ref{t:targets}. We show three of these targets in Figure \ref{f:targets}, together with the prediction/simulations made by each model. This figure also serves to illustrate three distinct predictive situations: the first target (panel a) is already well handled by the deterministic models, with visually adequate amount of detail and well-adjusted values. The second target (panel b) contains extreme values of precipitation, above 250 mm, which are not reached by deterministic approaches. The last target (panel c) illustrates cases in which deterministic models fall short, producing predictions that are \textit{smoothed out} with respect to their reference. Note that, this figure also illustrates the spatial \textit{fragmentation} in simulations made by the B-Gamma model. We have also analyzed other targets, but they were left out of the main article to focus on the most representative. For example, extreme dry targets (targets with no wet gridpoints) are overall very well reproduced by all models: very precisely by the deterministic model and with almost no variability for generative models.

To assess these targets we use a combination of metrics and figures. Note that, to fully explore the distributions produced by the generative models, all these metrics are computed from 1000 simulations drawn for the large-scale predictors corresponding to each target. First, we compute continuous ranked probability scores (CRPS, \citeA{gneiting_strictly_2007}). CRPS is a proper scoring rule that measures the discrepancy between the predictive distribution and the observed value. CRPS is expressed in the same units as the reference variable (here precipitation in millimeters per day), and collapses to the mean absolute error (MAE) if the prediction is deterministic, so it allows for a comparison with point predictions. Together with the CRPS maps, we include \textit{out-of-envelope error} maps and SAL (structure-amplitude-location) diagrams. 

The out-of-envelope error maps illustrate if the target values are beyond or inside the distributional limits of the generative models, and how far (above or below) in mm if they do so. That is, for the RCM-simulated (reference) precipitation value for a specific gridpoint $y$, we compute, for a set of simulations $S_m$ produced by a model $m$ for a specific day,  $$\left\{
\begin{array}{ll}
y - \max_{\hat{y} \in S_m}(\hat{y}) & \text{if } \  y  >  \max_{\hat{y} \in S_m}(\hat{y}), \\[0.4em]
y - \min_{\hat{y} \in S_m}(\hat{y}) & \text{if } \ y  < \min_{\hat{y} \in S_m}(\hat{y}), \\[0.4em]
0 & \text{if } y \  \in 
\left[
\min_{\hat{y} \in S_m}(\hat{y}),
\max_{\hat{y} \in S_m}(\hat{y})
\right].
\end{array}
\right.$$
A 0 in this metric means that the target reference value is within the distributional limits produced by the generative models. As for the climatological metrics, we also plot the out-of-envelope error maps with prior 3x3 moving averages, which allows us to overcome the limitation of gridpoint-based statistics. Indeed, this spatial \textit{smoothing} makes the diagnostic less sensitive to isolated gridpoint mismatches and more focused on whether the models cover coherent precipitation structures at a local spatial scale. 

The final assessment is done by means of the SAL diagrams \cite{wernli_sal_2008}. The SAL diagrams aim to evaluate how close are two precipitation fields when looking at the characteristics of so-called precipitation \textit{objects}, beyond fixed gridpoint or local statistics. Broadly speaking, precipitation objects are continuous fields of precipitation over a specific threshold within a domain. The threshold is set as suggested in the follow-up article \citeA{wernli_spatial_2009} and as in \citeA{doury_suitability_2024}: at $1/15$th of spatial quantile 95. A SAL score, shown as a 3-dimensional diagram, shows three metrics indicating properties of these precipitation objects, in essence (check \citeA{wernli_sal_2008} for a detailed mathematical description): structure, evaluating their size and shape; amplitude, their average precipitation amounts; and location, evaluating their geographical locations. If all objects are similar to the reference RCM-simulated ones, the three components will be close to 0. The SAL diagrams are used in the present study to assess the shape of the distribution of objects within a precipitation field, as generated by diffusion approaches, and compared against the objects generated by the deterministic model.

\begin{center}
    
\begin{sidewaystable}
{

\begin{tabular}{ |c|l|c|c|  }
\hline
 \textbf{Name} & \textbf{Description} & \textbf{Figure} & \textbf{Type} \\ \hline
 RMSE & Root mean squared error (predicted expected value vs reference). & \ref{f:rmse} & Climatological  \\  
 Mean & Relative bias (\%) in mean precipitation on wet days ($>1mm$). & \ref{f:mean_dry} & Climatological  \\
 Dry\% & Relative bias (\%) in the percentage of dry days ($\leq1mm$). & \ref{f:mean_dry} & Climatological    \\
 SD &  Relative bias (\%) in the standard deviation of the rainfall distribution.  & \ref{f:p01_sd_p99} & Climatological  \\
  Median & Relative bias (\%) in the median of the rainfall distribution. & \ref{f:p01_sd_p99}  & Climatological  \\
 P99 &  Relative bias (\%) in the 99th percentile of the rainfall distribution.  & \ref{f:p01_sd_p99} & Climatological    \\
 QQWet\% &  Quantile-Quantile plot of the spatial percentage of wet gridpoints. & \ref{f:qqplots} & Climatological   \\
 QQSpatial &  Quantile-Quantile plot of the spatially averaged rainfall. & \ref{f:qqplots} & Climatological \\
 RAPSD & Climatology of the radially averaged power spectral density. & \ref{f:raspd_climatology} & Climatological \\
  Out-of-envelope Error & Maps of differences between the minimum/maximum simulated and the observed value. & \ref{f:out-of-envelope error}  & Targeted  \\
 CRPS/MAE & Continuous ranked probability score (deterministic: MAE) for the predicted distribution. & \ref{f:crps_targeted} & Targeted \\ 
 SAL  & Structure-Amplitude-Location diagrams. & \ref{f:sal} & Targeted  \\
      \hline
\end{tabular}
}
\vspace{0.3cm}
\caption{Summary of the different evaluation metrics employed throughout the article. The column name displays the codename given in this article. Rainfall refers to the positive values of the precipitation distribution (over $1$ mm). Description provides a summary of the metric. Figure indicates in which figure they are displayed; Type indicates whether the metric is computed as a statistic over the whole evaluation period (\textit{Climatological}, Section \ref{s:climatological_analysis}), or for specific targets (\textit{Targeted}, Section \ref{s:targeted_analysis})}\label{t:metrics}
\end{sidewaystable}
\end{center}

\begin{center}
    
\begin{sidewaystable}
{

\begin{tabular}{ |c|l|c|c|c|c|c|  }
\hline
 \textbf{Name} & \textbf{Description} & \textbf{Date} & \textbf{Mean} & \textbf{Max} & \textbf{\% Wet} & \textbf{SD} \\ \hline
 Wettest & Day with the highest total precipitation over the whole domain. & 18/09/2071 & 24.4 & 172.4 & 77.8  & 27.5 \\  
 Max \% Wet  & Day with the highest number of wet gridpoints ($>1mm$). & 04/12/2009 & 14.4 & 47.6 & 100 & 6.7\\
 Wettest Gridpoint & Day with the maximum precipitation at any gridpoint.  & 15/10/2098 & 4.8 & 268.5 & 27.1 & 20.8   \\
 Highest SD & Day with the highest spatial variance.  & 11/10/2052 & 18.2 & 242.6 & 73.2 & 33.2 \\
  Drizzle & Lowest precipitation amount divided by \% of wet gridpoints. & 13/12/2008  & 1 & 6.9 & 53.5 & 0.9 \\
 Historical & Selected from \cite{doury_suitability_2024}.  & 25/12/2022 &  2.2 & 30.3 & 46.6 & 3.4 \\
 Mid-future & Selected from \cite{doury_suitability_2024}.  & 10/07/2045 &  4.9 & 55.8 & 57.9 & 7.2 \\
 End-Future & Selected from \cite{doury_suitability_2024}.  & 05/10/2099 & 14.2 & 141.5 & 80.3 & 15.3 \\ \hline
\end{tabular}
}
\vspace{0.3cm}
\caption{The eight target precipitation fields, taken from the evaluation period considered in this study (RCP4.5, 2005-2100) for the in-depth analysis in Section \ref{s:targeted_analysis}. Note, \textit{Max \% Wet} was selected randomly among the targets that fulfill the condition that $100\%$ of the spatial domain is wet. The columns Mean, Max, \% Wet and SD indicate the spatial statistics (in mm) of each target day, respectively: mean precipitation, maximum precipitation, percentage of gridpoints over $1$mm, and spatial standard deviation.}\label{t:targets}
\end{sidewaystable}
\end{center}

\section{Results}\label{s:results}

This section is divided into two parts. In Section \ref{s:climatological_analysis}, we perform a climatological evaluation of the models considered. In Section \ref{s:targeted_analysis}, we analyse the uncertainty envelopes produced for selected target events (Table \ref{t:targets}) and assess their added value.

Recall that all models are trained using data from the Historical and RCP8.5 scenarios, covering the combined period 1951--2100, and are then applied to predictors from the RCP4.5 scenario over the period 2005--2100. Therefore, all evaluations, including the selected target events, correspond to simulations and predictions performed for this independent evaluation period. We therefore treat the RCM-simulated statistics and values as the truth against which the models are evaluated, and refer to them as the \textit{reference} throughout the text and figures.

We also refer to models capable of producing stochastic simulations (B-Gamma, CPMGEM, and ParamDiffusion) as \textit{generative}, for simplicity. Unless explicitly stated otherwise, generative models are assessed by sampling from them: 50 simulations are used for the full evaluation period in Section \ref{s:climatological_analysis}, whereas 1000 simulations are drawn from each generative model for each target event in Section \ref{s:targeted_analysis}.

\subsection{Climatological Analysis}\label{s:climatological_analysis}
In this section we perform a climatological analysis of the models considered. We recall that, here, climatology refers to a statistic measured over the whole evaluation period. 

In terms of predictive performance, as measured by the RMSE, the average (spatially) RMSE is $3.5, 3.48, 3.18$, and $3.29$ respectively, for Asym, B-Gamma, CPMGEM, and ParamDiffusion. Asym and B-Gamma perform almost equally, with both diffusion approaches not only keeping up with the performance, but also slightly improving it: on average $\sim 0.3$ ($\sim 0.2$) mm with respect to ParamDiffusion (CPMGEM). With respect to both diffusion models, CPMGEM performs slightly better than ParamDiffusion (on average 0.11 mm). Considering the difference in the number of parameters and cost in ParamDiffusion (approximately one quarter of the simulation cost), we think this is a good compromise. This is also true for Asym and B-Gamma, which have around $2.5\%$ the total number of trainable parameters of the backbone of ParamDiffusion and around $0.6 \%$ those of CPMGEM, and allow for simulations/predictions almost instantly. The backbone of both diffusion approaches also have additional more complex mechanisms such as attention, which could potentially improve this result, but this is beyond the scope of this article.

\begin{figure}[tb]
\centering 
  \makebox[0pt][c]{%
    \includegraphics[width=1.2\linewidth]{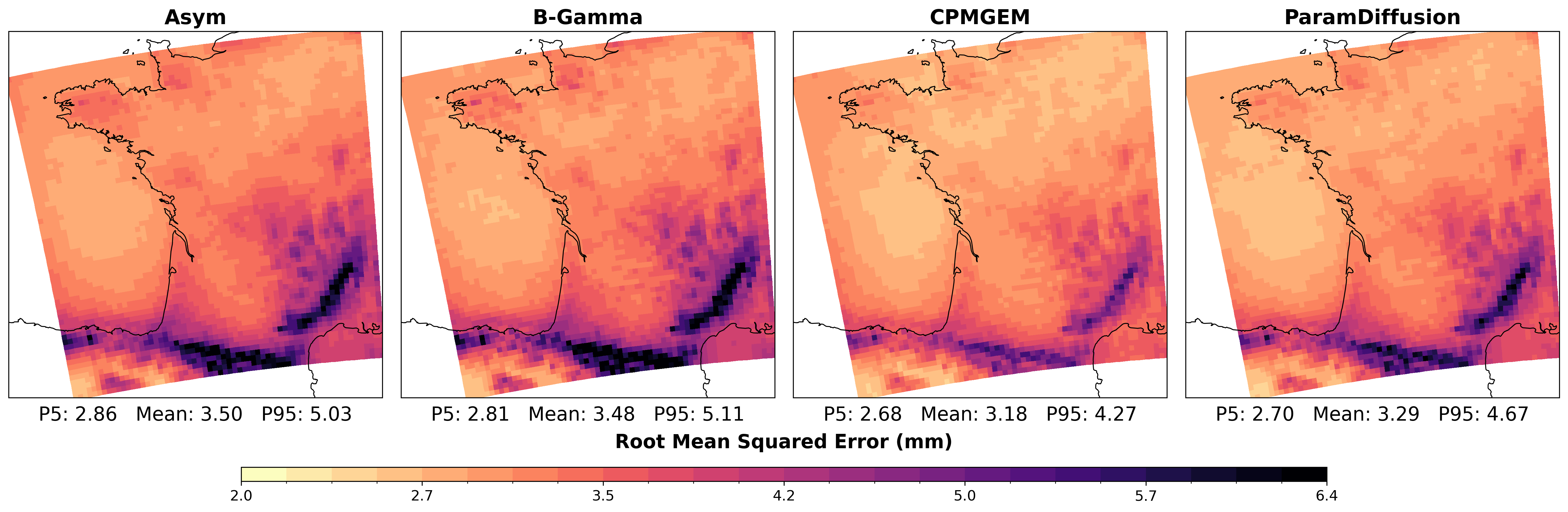}
  }

  \caption{Prediction error as measured by the root mean squared error (RMSE), computed for the whole evaluation period, for the different models considered. In the case of the generative models, the value used for the computation is the expected value for each day, as the average of 50 simulations.}
  \label{f:rmse}
\end{figure}

\begin{figure}[tb]
\centering
  \makebox[0pt][c]{%
    \includegraphics[width=1.2\linewidth]{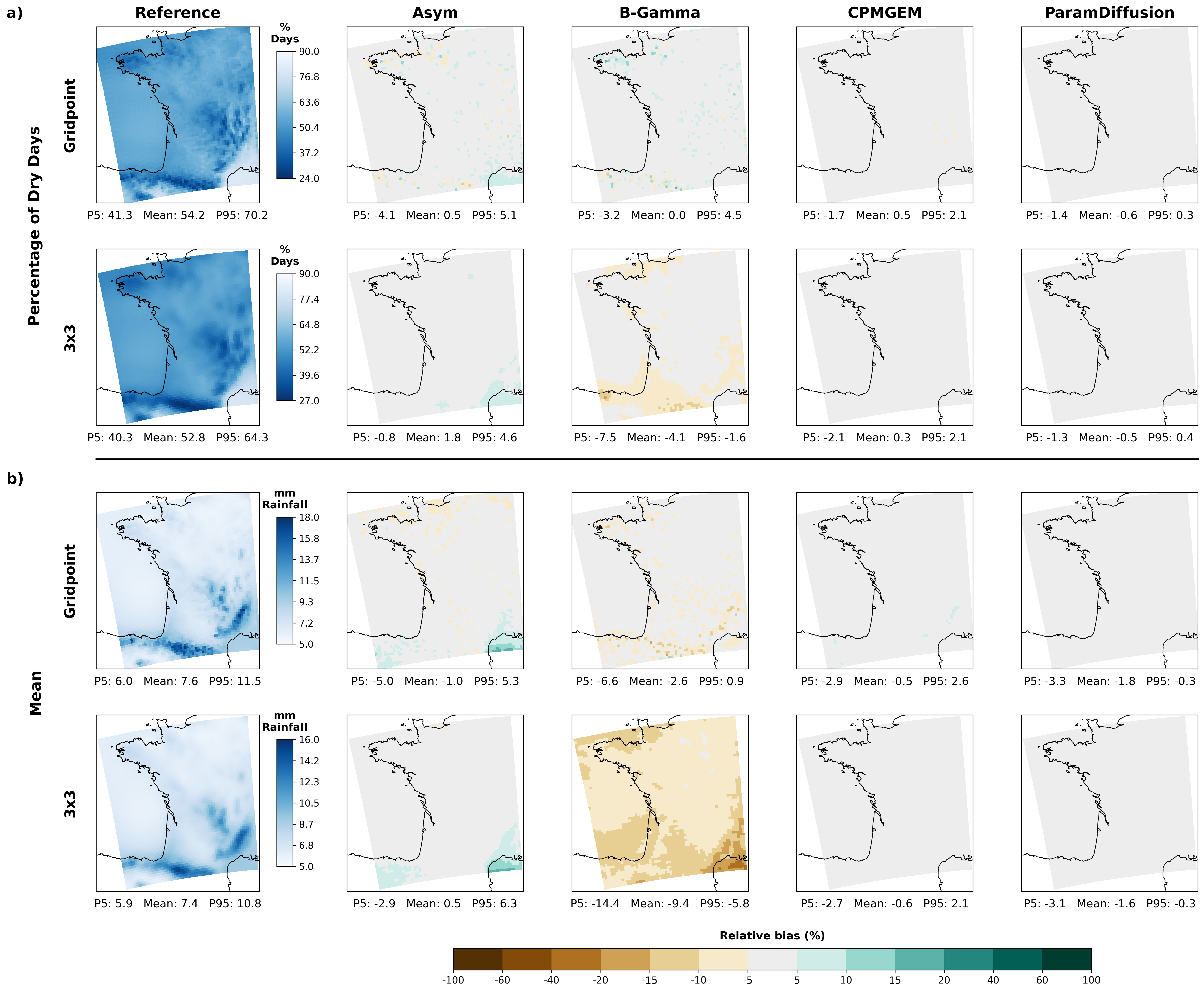}
  }
  \caption{Relative bias in the percentage of dry days (panel a) and  mean rainfall over wet days (1mm, panel b) for the models considered (columns). The reference climatology is shown on the first column. For both metrics, results are show per gridpoint (top row, for each panel), and using a 3x3 moving average window (bottom) before computing the statistics, to account for the local spatial consistency.}
  \label{f:mean_dry}
\end{figure}

\begin{figure}[tb]
\centering
  \makebox[0pt][c]{%
    \includegraphics[width=1.2\linewidth]{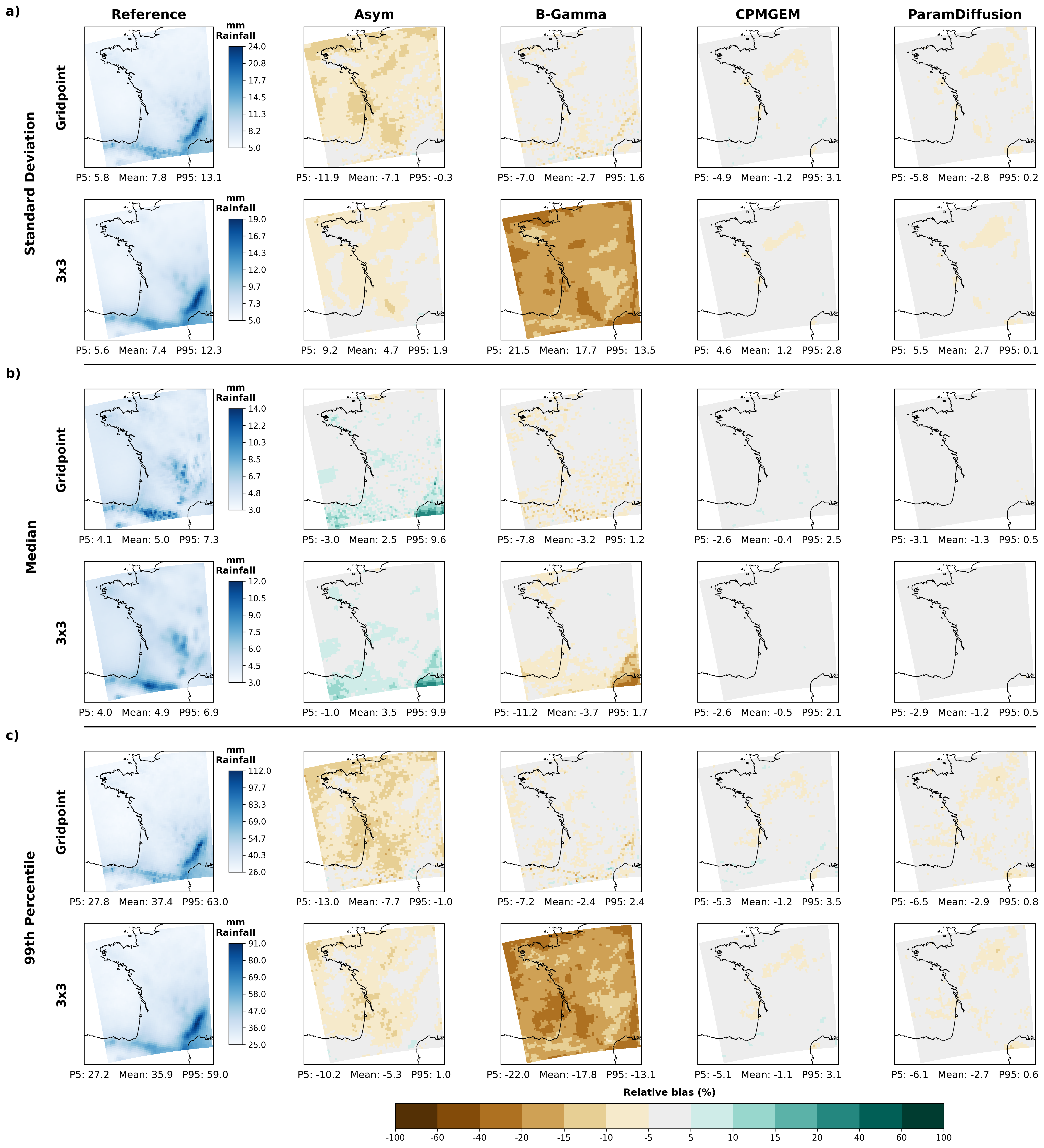}
  }
  \caption{As Figure \ref{f:mean_dry}, for the standard deviation (panel a), Median (panel b) and 99th percentile (panel c); for values over 1mm.}
  \label{f:p01_sd_p99}
\end{figure}

Biases at the gridpoint level and in spatial distributions over the whole evaluation period are shown in Figures \ref{f:mean_dry} and \ref{f:p01_sd_p99}. When it comes to the percentage of dry days (\%Dry) and mean precipitation over wet days (Mean) in Figure \ref{f:mean_dry}, all models perform well. This is both for the gridpoint statistics and the local spatial ones (i.e., 3x3 statistics), with the exception of the B-Gamma model, which fails to capture the spatial means, especially for the average rainfall, and exhibits a dry bias of around $10\%$ on average. Some minor wet and dry biases are observed for Asym and B-Gamma, respectively, while both diffusion approaches reproduce these statistics, including their spatial aspects, realistically.

Biases on more complex statistics of the distribution are clearly more difficult to capture for all models. Looking at all the panels for both diffusion models together, it is clear that they excel at capturing distributional aspects, as has been reported in other studies. Interestingly, SD and P99 biases for the Asym model are reduced slightly when looking at 3x3 maps. This is not the case for the Median, for which biases are slightly higher. The combination of dry and wet biases present in Asym indicates biases in the shape of the distribution, rather than a consistent bias. Nevertheless, for Asym, biases in the most ambitious climatological statistic assessed in this study, P99, are relatively contained, with 5th spatial percentile being $-13 \%$. Contrary to what many studies report, this highlights how deterministic models trained to optimize a loss function well adapted for precipitation, can capture the tail of the distribution reasonably well. Considering the substantially lower training time and the almost immediate prediction time (compared to the hours or days needed for the diffusion models, see Table \ref{t:models}), this should be taken into account.
Finally, the spatial pixelation effect observed already for the B-Gamma model is again present here, for the SD and P99.

\begin{figure}[tb]
\centering
  \makebox[0pt][c]{%
    \includegraphics[width=1.2\linewidth]{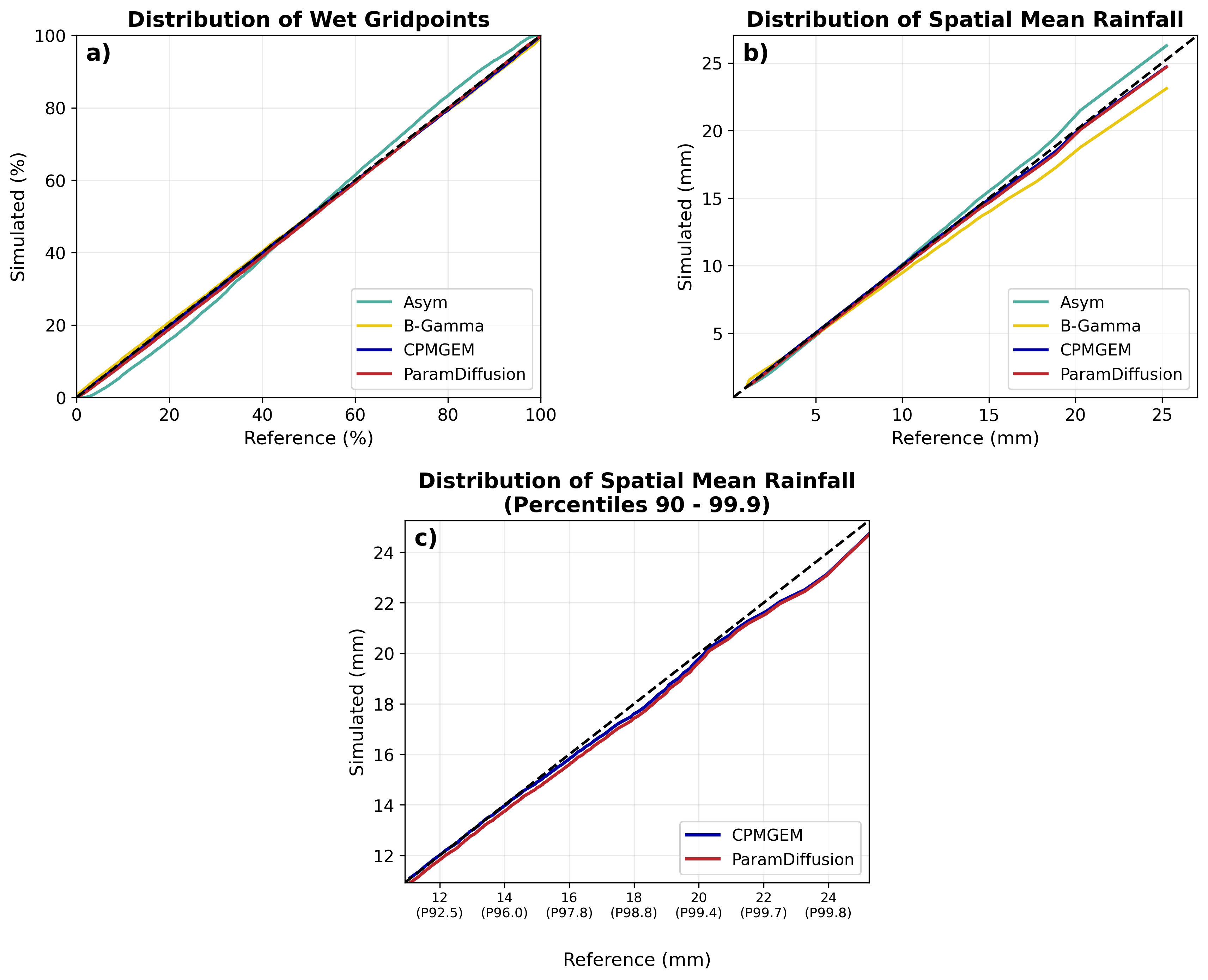}
  }
  \caption{Comparison of the spatial distributions (for the whole domain), using quantile-quantile plots. Panel (a) shows the percentage of wet days and panel (b) shows the average rainfall over 1mm for the whole domain. Panel (c) shows the average rainfall for days between percentiles 90 to 99.9. Only the 2 diffusion models are shown in this panel.}
  \label{f:qqplots}
\end{figure}

In Figure \ref{f:qqplots}, we show the domain averaged distributions of wet gridpoints, as percentage, and spatial rainfall on wet days (over 1 mm). The distribution of wet gridpoints is captured by the three generative models, with Asym exhibiting some moderate bias in the shape of the distribution. When it comes to the average rainfall, the four models perform well for the lower half of the distribution, with Asym overestimating domain-wide higher percentiles and B-Gamma underestimating them. The overestimation effect seen in Asym likely comes from the cost function aiming at higher values of the gridpoints' distribution, with the bias for B-Gamma coming from the spatial fragmentation, that renders the model incapable of capturing domain wide precipitation. Both diffusion approaches generate very realistic distributions up to very high quantiles, so, in panel (c), we have zoomed in on the tail and assess quantiles up to $0.999$ probability on wet days. Both diffusion models have very similar performance and slightly underestimate the distribution by a maximum of 1 mm of precipitation in quantiles beyond $97\%$.

\begin{figure}[tb]
\centering
  \makebox[0pt][c]{%
    \includegraphics[width=1.2\linewidth]{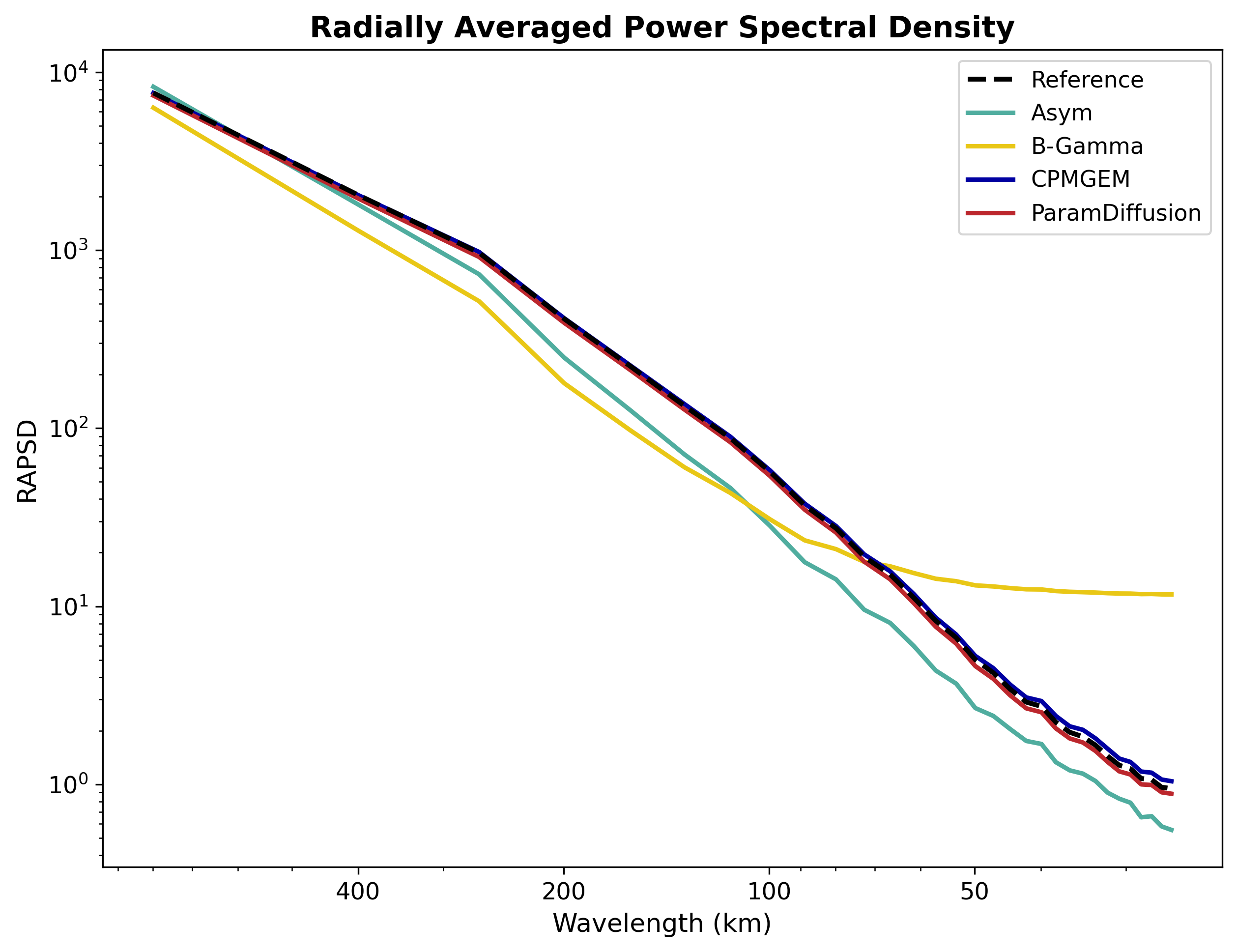}
  }
  \caption{Radially averaged power spectral density (RAPSD) for the models considered in this study for the deterministic model (Asym, teal) and  simulations from B-Gamma, yellow; CPMGEM, dark blue; and ParamDiffusion, red. The RAPSD for the reference RCM-simulated precipitation fields is shown as a dashed black line.}
  \label{f:raspd_climatology}
\end{figure}

Finally, to conclude this climatological analysis, we address the RAPSD in Figure \ref{f:raspd_climatology}, in order to compare the variability over different spatial scales produced by different approaches, thus indicating whether models produce the right amount of detail as compared to the reference RCM-simulated precipitation fields. Asym falls behind when representing higher frequency details (beyond 300km), with B-Gamma simulations showing inconsistent RAPSD values due to the spatial fragmentation. Diffusion-based approaches produce very realistic detail levels at all scales, which is consistent with previous studies \cite{addison_machine_2026, mardani_residual_2025}.

Overall, notably, both diffusion approaches perform similarly in all climatological aspects considered. Considering the substantial differences between both approaches, this leads us to conclude that this is a systematic result, rather than a model-dependent feature.

\subsection{Targeted Uncertainty Envelope Analysis}\label{s:targeted_analysis}

As explained in Section \ref{ss:evaluation_methods}, the climatological analysis is insufficient for fully understanding how generative approaches perform when faced by specific extreme situations. The previous section showed that diffusion models faithfully reproduce the climatologies, including their spatial counterparts. However, the extent of the added value of the additional spatial details produced remains to be properly assessed. Then, in this section, we focus on the target RCM-simulated precipitation fields selected in Section \ref{ss:evaluation_methods} and detailed in Table \ref{t:targets}. We recall that the most interesting feature of a generative emulator is perhaps their potential to generate climate extremes consistent with the large-scale predictors' state. In this cases deterministic models have to compromise on a specific fixed field, which, by definition, will not be an extreme.

\begin{figure}[!htb]
\centering
  \makebox[0pt][c]{%
    \includegraphics[width=1.2\linewidth]{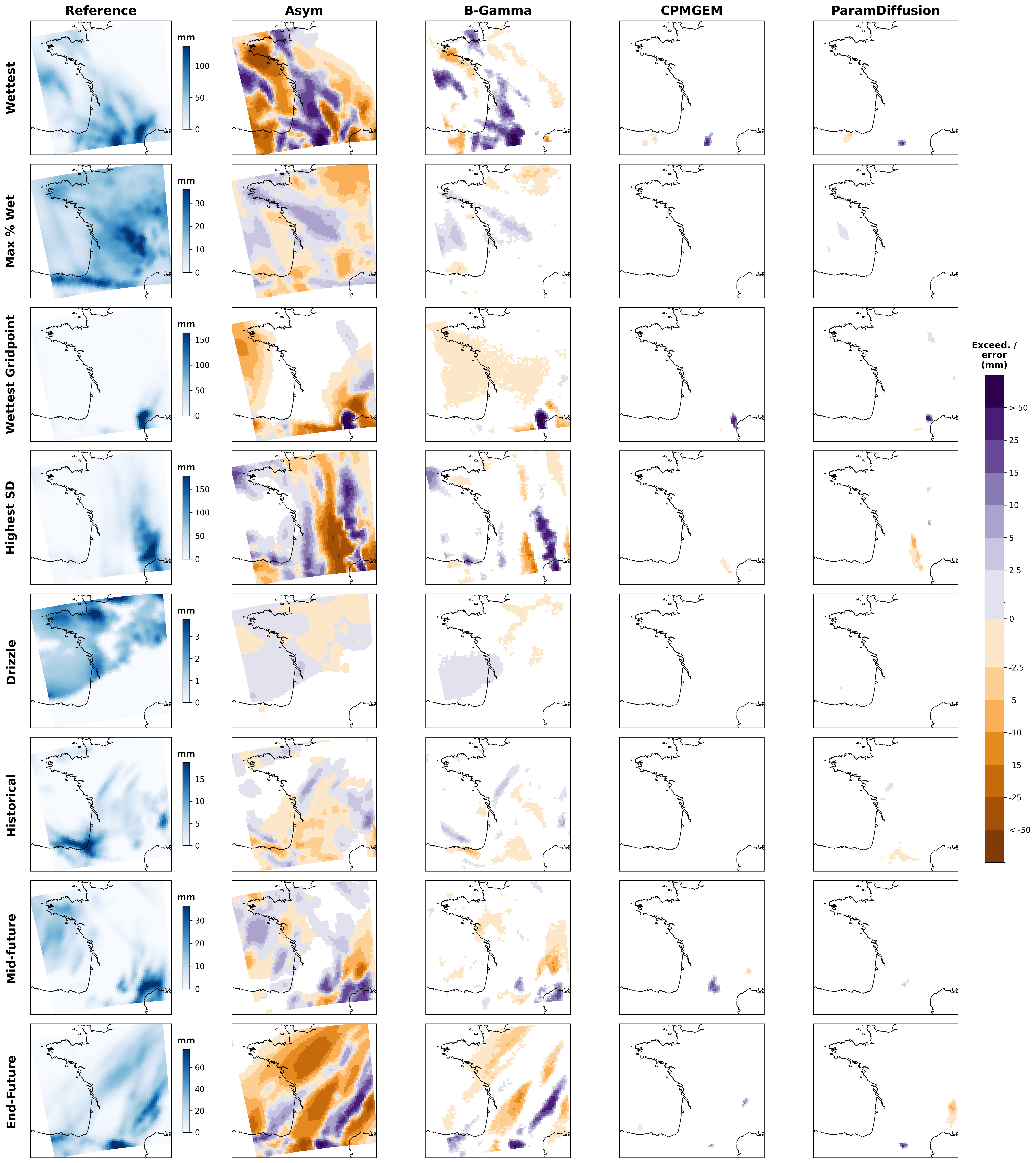}
  }
  \caption{For each selected target day (rows), out-of-envelope error for each model (columns, error for Asym), with a prior 3x3 moving window. For 1000 simulations, out-of-envelope error is computed as the positive/negative difference between the maximum/minimum value generated by the generative model vs the observation (purple/brown); and is 0 (white) if the observed value lies within the simulated distribution limits.}
\label{f:out-of-envelope error}
\end{figure}

\begin{figure}[!htb]
\centering

  \makebox[0pt][c]{%
    \includegraphics[width=1.2\linewidth]{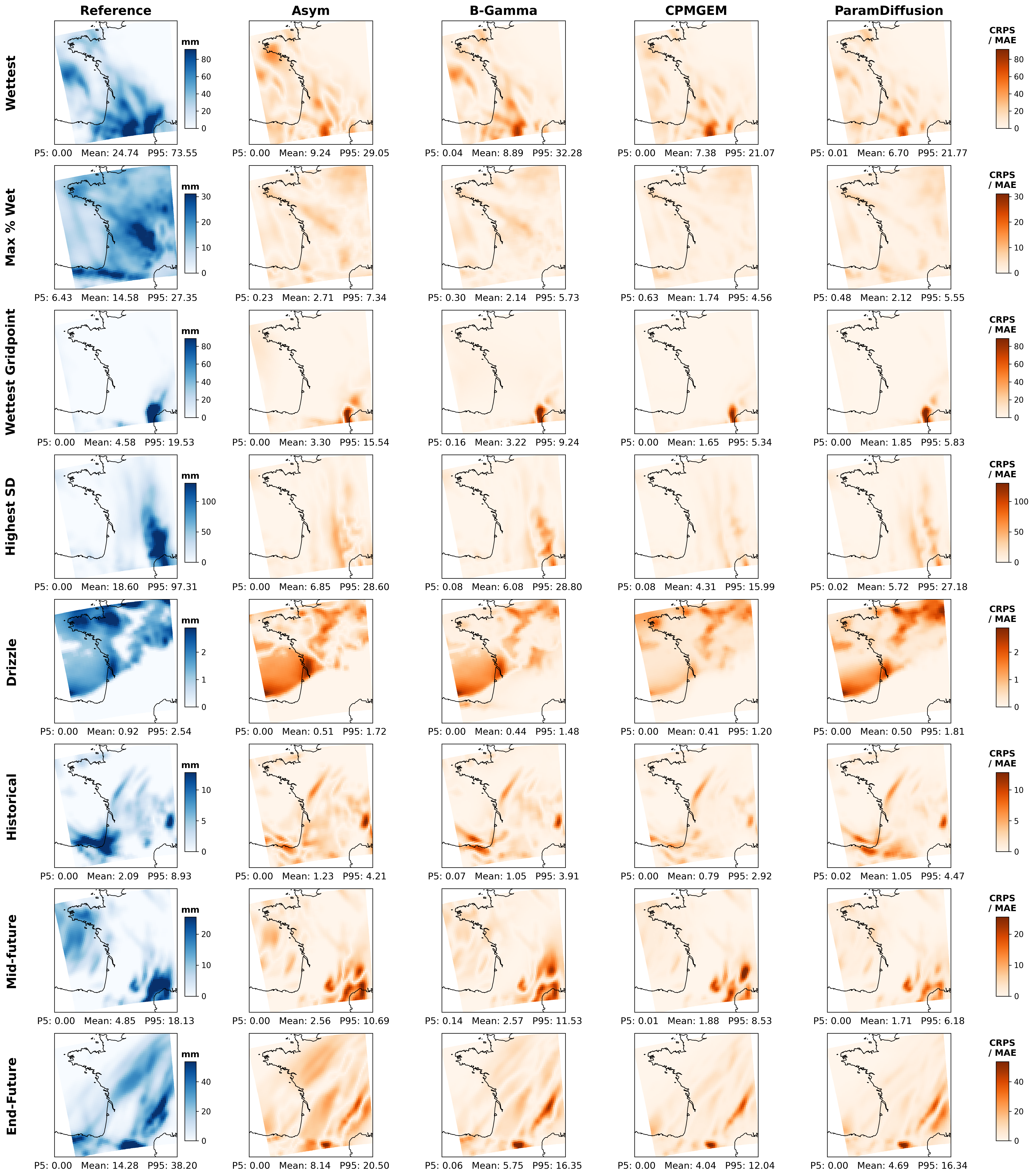}
  }
  \caption{For the selected target reference RCM-simulated precipitation fields (rows, see Table \ref{t:targets}), continuous rank probability scores (CRPS) for the models considered (columns), with a previous 3x3 moving average window. Note that, for the deterministic model (Asym) this corresponds to the mean absolute error.}
\label{f:crps_targeted}
\end{figure}

\begin{figure}[!htb]
\centering
  \makebox[0pt][c]{%
    \includegraphics[width=1.2\linewidth]{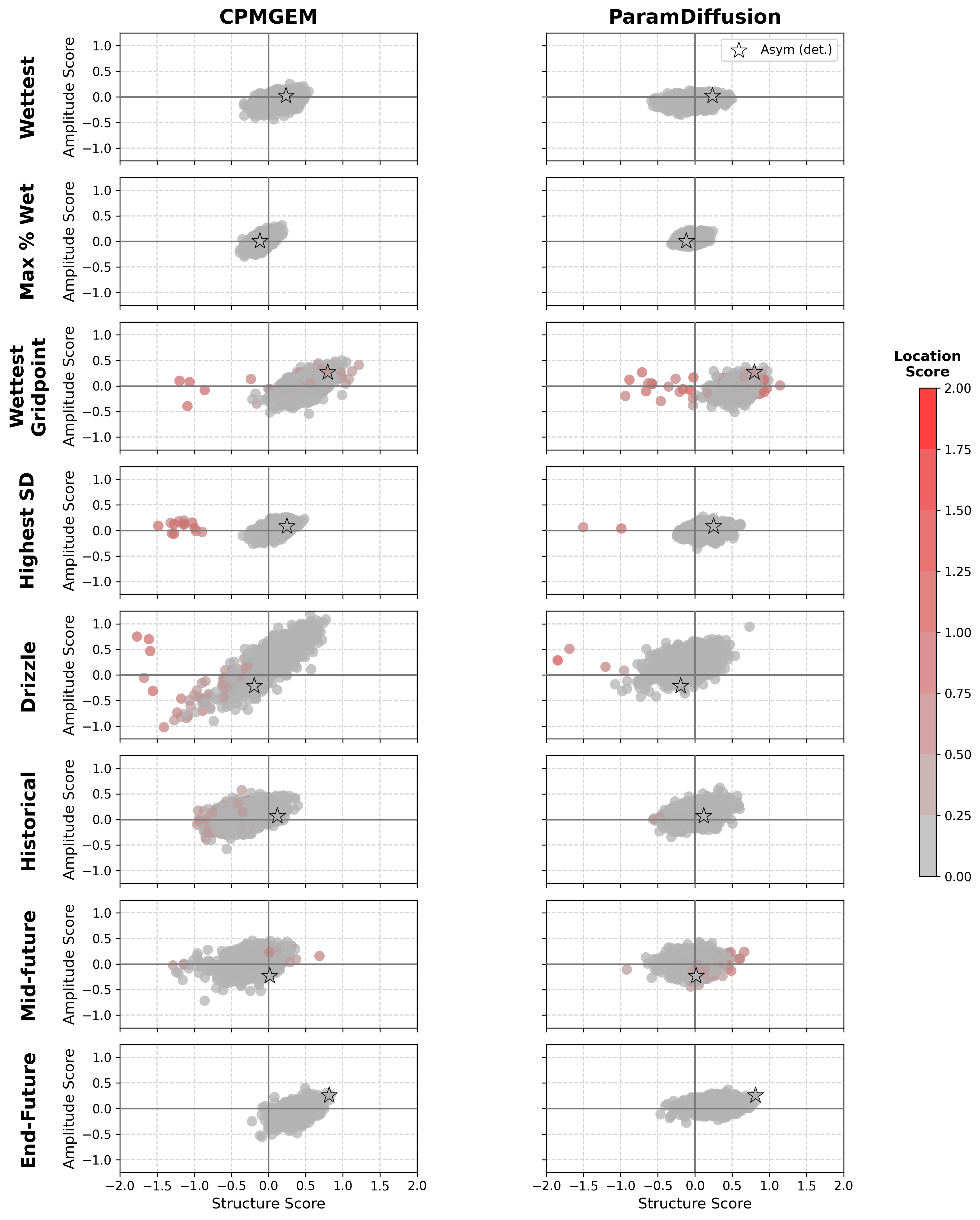}
  }
  \caption{SAL diagrams for 1000 simulations (dots) for each target (rows), the deterministic prediction made by Asym (star), CPMGEM (first column) and ParamDiffusion (second column). Note that the y-axis has been capped at $[-1,1]$ since there is no point outside this range.}
\label{f:sal}
\end{figure}

The analysis performed in this section approaches the problem from complementary angles. The out-of-envelope error maps in Figure \ref{f:out-of-envelope error} provide a simple way to assess whether the reference values fall within the distributional limits produced by each model; that is, whether such extreme instances, which are known to be possible under the corresponding large-scale conditions, can be simulated by the diffusion approaches. Complementing this analysis, Figure \ref{f:crps_targeted} shows the CRPS scores for the four models assessed in this work, while the SAL diagrams in Figure \ref{f:sal} provide a final event-based assessment of precipitation objects produced by the two diffusion models. By construction, deterministic models are not expected to reproduce the most extreme target fields exactly. Therefore, we consider that a generative model adds value if it fulfills the following three conditions: the uncertainty cloud covers the observed values (Figure \ref{f:out-of-envelope error}), its ensemble of simulated precipitation objects shifts the distribution of SAL scores towards the origin in the three dimensions considered (structure, amplitude, and location, Figure \ref{f:sal}), all while remaining controlled and avoiding unrealistic values, that is, preserving predictive skill as measured by CRPS (Figure \ref{f:crps_targeted}).

The out-of-envelope error maps in Figure \ref{f:out-of-envelope error}, which are shown after applying a 3×3 moving average prior to the computation (the same figure, without the 3x3 moving average, is provided in the supporting information file: Figure S1), show that both CPMGEM and ParamDiffusion generally produce uncertainty envelopes that include the target precipitation fields over most of the domain:  out-of-envelope errors are usually sparse and spatially limited. For targets such as Max  \% Wet, Drizzle, and Historical, the reference fields are almost entirely contained within the simulated ranges. In these cases, the observed RCM field can therefore be interpreted as a plausible realization of the conditional distribution learned by the diffusion models. Note that these three targets exhibit comparatively low precipitation values and that, generally,  they are already accurately predicted by deterministic approaches.

The most challenging cases are the targets with intense and localized precipitation structures, notably Wettest Gridpoint. For Wettest, Wettest Gridpoint, Highest SD, and End-Future, some out-of-envelope errors can be seen for the highest-intensity areas. These out-of-envelope errors indicate that the diffusion models do not always fully cover the upper tail of the conditional precipitation distribution, especially when extreme values are spatially concentrated. Depending on the target, each diffusion approach behaves differently: for instance, for Mid-future, CPMGEM fails to capture in its uncertainty envelope a precipitation object in the south of the domain, which is not the case of ParamDiffusion. Conversely, in other cases, such as Highest SD, we see that ParamDiffusion produces a distribution that is too wet, thus observing some negative out-of-envelope errors in the southeast of the domain. In any case, out-of-envelope errors remain much more localized for both diffusion approaches if we compare them to the error for deterministic reference, with diffusion models, as expected, representing a wider range of outcomes, even though without reaching the extreme values observed. For example, in the case of Wettest Gridpoint, we observe that the maximum error made by Asym is 148.7 mm, clearly falling short to predict the extreme storm in the southeast. For the two diffusion models, this is greatly reduced: For CPMGEM these values are 54.9 mm (maximum) and 21.4 mm (on average, spatially); whereas for ParamDiffusion, the best scoring model for this specific case, they are 50.6 mm (maximum) and 10.5 mm (on average, spatially).

We observe also how the B-Gamma model presents incoherent distributions, consistent with the qualitative analysis in Figure \ref{f:targets}. Note, however, that B-Gamma produces the most realistic uncertainty envelope in terms of out-of-envelope error in a gridpoint-based analysis (see Figure S1 in the supporting information file), which validates its use for gridpoint analyses and as background model for ParamDiffusion. 

The CRPS maps in Figure \ref{f:crps_targeted} confirm that, naturally, the largest probabilistic errors occur for the most intense targets. Wettest, Highest SD, and End-Future show the highest errors across models, with maxima attained in the same regions where the reference field contains the strongest precipitation. This confirms that a challenge remains to reproduce the conditional intensity distribution of extreme events. Across targets, the diffusion models improve over the deterministic reference in all cases. This improvement is particularly clear for the most extreme cases, where the spatial mean CRPS of CPMGEM and ParamDiffusion is substantially lower than the MAE of Asym. ParamDiffusion reaches the lowest spatial mean score for Wettest and Mid-future, while CPMGEM performs slightly better for some localized targets such as Max  \% Wet and Wettest Gridpoint. We confirm that the differences between the two diffusion models are therefore event-dependent rather than systematic. ParamDiffusion appears somewhat more controlled in several cases, whereas CPMGEM sometimes allows a broader range of local outcomes. Overall, we consider that no model ranks best. Note that we also provide the analogous to Figure \ref{f:crps_targeted}, without the 3x3 moving average, in the supporting information file: Figure S2.

Finally, we analyze the SAL diagrams in Figure \ref{f:sal}. For Wettest and Max \% Wet (i.e., first two rows), interestingly, the deterministic reference (Asym) already behaves well and very close to the origin, and both CPMGEM and ParamDiffusion generate relatively compact ensembles that stay close to the origin. This is desirable, since we consider that, for targets already well captured by deterministic models, the simulated ensemble should remain close to the origin rather than introducing unnecessary dispersion or degradation. Moreover, for the last target (End-future), where Asym performs poorly, we observe a behavior indicative of added value for diffusion models: both diffusion approaches produce an ensemble of simulations that clearly tilts towards the origin, that is, towards realistic precipitation objects. Notably, this happens for the structure score, which confirms, mathematically, the \textit{visual} improvement in the shape and detail we see for this target in Figure \ref{f:targets} (panel c).

In the case of Wettest Gridpoint, a very localized extreme, the cloud is much more dispersed and moves towards the origin, although some simulations show larger location errors, indicating that intense precipitation objects can be generated but are sometimes placed in an inadequate location (specially for ParamDiffusion), confirming that the models are not perfectly capturing the full uncertainty, as we saw in the previous figures. This is consistent with the CRPS maps, where errors remain concentrated around high-intensity structures, and with the out-of-envelope error maps, where we see errors localized around the strongest precipitation areas. The target Highest SD is halfway: the deterministic model already works relatively well in this case, and the distribution produced by diffusion models stays close to the origin, but some simulations exhibit unjustifiable both structure and location scores, notably for CPMGEM. The Drizzle case highlights a different limitation: although absolute precipitation amounts are low and CRPS values remain small, the SAL diagrams show a relatively broad spread in structure and amplitude. We clearly see that low-intensity precipitation objects are easy to capture by the deterministic models, with diffusion models (see e.g.,  panel a) in Figure \ref{f:targets}) generating a distribution of samples too wide and not necessarily consistent with the large-scale predictors. The uncertainty clouds produced for  Historical and Mid-future are more centered around the origin for ParamDiffusion, which is consistent with the analysis performed in the previous two figures, where CPMGEM exhibited higher out-of-envelope error values.

\section{Conclusions and Discussion}  
This study provides a comprehensive assessment of diffusion-based approaches for precipitation emulation in a perfect-model framework. We evaluated whether generative models can add value beyond state-of-the-art deterministic emulators by producing meaningful uncertainty envelopes conditioned on large-scale atmospheric predictors. To this end, we intercompared a deterministic precipitation emulator specifically designed for capturing the tail of the precipitation distribution (Asym), a parametric Bernoulli–Gamma probabilistic model (B-Gamma), a direct conditional diffusion benchmark (CPMGEM), and our proposed two-stage ParamDiffusion framework. Beyond the model intercomparison, we make two specific contributions. First, we introduce ParamDiffusion as a 2-stage generative framework that separates the extraction of the predictable large-scale signal (including its uncertainty) from the generation of the spatially consistent conditional uncertainty envelope. Second, we also propose a comprehensive and more thorough validation framework for generative precipitation emulators, combining climatological diagnostics with a targeted event-based analysis of selected precipitation fields of special relevance to climate scientists. This targeted analysis is a central contribution of this work, since it directly assesses, in detail, whether the simulated uncertainty envelope contains plausible realizations of specific extreme RCM-simulated events, whether probabilistic errors are reduced, and whether the simulated precipitation objects are realistic in structure, amplitude, and location.

The first conclusion is that diffusion models provide added value when representing climatologies regardless of the approach considered. The deterministic component of the downscaling function is already well captured by the Asym model, which in practice remains highly competitive despite its much lower computational cost. Diffusion approaches moderately improve this metric, arguably due to a much more complex backbone. Indeed, both diffusion approaches reproduce the distributional properties of precipitation with high realism for the whole evaluation period, and this result is not restricted to pointwise gridpoint statistics. The same conclusion holds when the diagnostics are computed after applying a local spatial aggregation, showing that the diffusion models capture not only marginal precipitation distributions, but also the local spatial aspects of precipitation. The similarity in climatological performance, exhibited by both diffusion approaches and despite their substantial methodological differences, suggests that the high performance shown in this article is a property of diffusion-based precipitation emulation. This is a key difference with respect to parametric models like B-Gamma, which produce realistic gridpoint distributions but not spatially-consistent precipitation fields. 

With respect to the specific targets analyzed, we show that good climatological behavior and visually realistic spatial precipitation fields do not guarantee that the conditional upper tail of extreme events is correctly captured. For several moderate or already well-predicted targets, the diffusion models generate uncertainty envelopes that contain the reference RCM field over most of the domain, reduce probabilistic errors relative to the deterministic baseline (in terms of CRPS scores), and produce precipitation objects that remain close to the reference, as measured by the SAL scores. This confirms that diffusion models can provide meaningful event-scale uncertainty information. In these cases, the generated ensemble does not merely add variability, it represents plausible alternative realizations compatible with the large-scale atmospheric situation. However, the most intense and localized events remain challenging. For targets with extreme precipitation structures, the diffusion models substantially reduce the errors made by the deterministic emulator, but they do not fully solve the problem. These results highlight a central difficulty of stochastic precipitation emulation: generating visually realistic looking extremes is not enough. The extremes must also be generated with the correct intensity, spatial structure, and location for the specific large-scale atmospheric state. None of the models assessed here fully achieves this for the most severe targets. 

This limitation has direct implications for the interpretation of generative downscaling models. Diffusion models clearly improve the realism of simulated precipitation fields and provide more informative uncertainty estimates than deterministic approaches. Yet, they should not be considered reliable generators of the most severe precipitation extremes solely because they reproduce climatological tails or visually detailed fields. Standard diagnostics such as RMSE, mean bias, or RAPSD are therefore insufficient to validate generative climate emulators. They must be complemented by diagnostics that explicitly assess the conditional uncertainty envelope, as is done in this study. Thus, targeted event-based analyses provide a necessary counterpart and we argue that it should become a standard component of the evaluation of generative emulators, alongside traditional climatological diagnostics. 

Overall, our results show that diffusion models, no matter the approach, are powerful tools for generative RCM emulation, and capture reliably climatological uncertainty, generating spatially coherent ensembles that preserve distributional properties. Nevertheless, as said before, their main objective is not simply to add more small-scale texture or improve visual realism. Learning the conditional upper tail of precipitation extremes remains a challenge: when an extreme event is possible under a given large-scale atmospheric situation, the emulator should be able to generate it. This should be achieved without the model generating situations incompatible with the large-scale predictors' state, generally referred to in literature as \textit{hallucinations} \cite{aithal_understanding_2024}. Until this is achieved, diffusion-based emulators should be viewed as powerful tools for probabilistic climatological emulation, but not yet as fully reliable substitutes for RCMs in the simulation of the most severe localized precipitation extremes. Note that this should be taken into account considering that diffusion models are computationally expensive and require hours to generate a single member, versus the almost immediate generation that models such as Asym can achieve. This study also highlights how CPMGEM and ParamDiffusion perform very similarly across most climatological diagnostics, despite substantial differences in formulation, conditioning strategy, number of parameters, and computational cost.

Finally, we foresee several other lines of future work, including the study of hallucinations, mentioned before, and physical inconsistencies. We plan to also expand this work to assess other steps in the modeling chain and study how different models behave when transferred to GCM-simulated large-scale fields. Also, the reduction of computational costs and expanding this study to other generative methods (such as flow matching methods, see \citeA{albergo_building_2023, lipman_flow_2023}) is an important research perspective. We note that, although this work has focused on precipitation, the same modeling strategy is not restricted to this variable. Preliminary experiments with maximum daily temperature, as well as applications to other geographical domains, showed very promising performance. This suggests that the framework generalizes well to other climate variables and regions and will be addressed in future work.

\section*{Open Research Section}
All the code necessary to reproduce the results has been made available on GitHub: 
 \url{https://github.com/MNLR/Diffusion}. Similarly, all the data necessary to reproduce the results can be found on Zenodo: \url{https://doi.org/10.5281/zenodo.20610814}.

\section*{Conflict of Interest declaration}
The authors declare there are no conflicts of interest for this manuscript.

\acknowledgments
This study has received funding managed by the \textit{Agence Nationale de la Recherche} under France 2030 bearing the references ANR-22-EXTR-0005 (TRACCS-PC4-EXTENDING project) and ANR-22-EXTR-0011 (TRACCS-PC10-LOCALISING project). This project was provided with computing AI and storage resources by GENCI at IDRIS thanks to the grant 2025-AD010116371 on the supercomputer Jean Zay's V100 partition. It also benefited from the IPSL Data and Computing Center ESPRI which is supported by CNRS, SU, CNES and Ecole Polytechnique.

%
%

\bibliography{main}

%
%
%
%
%

\end{document}